\documentstyle[aps,prl,preprint]{revtex} 

\tighten
\begin{document}
\draft

\title{Quantum repeaters for communication}

\author{
H.--J.\ Briegel$^{1,2,*}$, W. D\"ur$^1$,
J. I. Cirac$^{1,2}$, and P. Zoller$^1$}
\address{$^{1}$Institut f\"ur Theoretische Physik, Universit\"at Innsbruck,
Technikerstrasse 25, A--6020 Innsbruck, Austria.\\
$^{2}$Departamento de Fisica, Universidad de Castilla-La Mancha,
13071 Ciudad Real, Spain.}

\date{\today}

\maketitle

\begin{abstract}
In quantum communication via noisy channels, the error probability
scales exponentially with the length of the channel. We present a scheme
of a quantum repeater that overcomes this limitation. The central idea
is to connect a string of (imperfect) entangled pairs of particles by
using a novel nested purification protocol, thereby creating a single
distant pair of high fidelity. The scheme tolerates general errors on
the percent level. 
\end{abstract}



\narrowtext


Quantum communication deals with the transmission and exchange of
quantum information between distant nodes of a network. Remarkable
experimental progress has been reported
recently, for example, on secret key distribution for quantum
cryptography \cite{tittel97,buttler97}, teleportation of the
polarization state of a single photon \cite{bouwmeester97,boschi98}, and
the creation of entanglement between different atoms \cite{hagley97}. On
the other hand, first steps towards the implementation of quantum
logical operations, which are the building blocks of quantum computing,
have been demonstrated \cite{monroe95_turchette95}. In view of this
progress, it is not far-fetched to expect the creation of small quantum
networks in the near future. Such networks will involve nodes, where
qubits are stored and locally manipulated, and which are connected
by quantum channels over which communication takes place by sending
qubits. This will open the possibility for more complex activities such
as multi-party communication and distributed quantum computing
\cite{grover97}. 

The bottleneck for communication between distant nodes is the scaling of
the error probability with the length of the channel connecting the
nodes. For channels such as an optical fiber, the probability for both
absorption and depolarization of a photon (i.e. the qubit) 
grows exponentially with the length $l$ of the fiber. 
This has two effects: $(i)$ to transmit
a photon without absorption, the number of trials scales exponentially
with $l$; $(ii)$ even when a photon arrives, the fidelity of the
transmitted state decreases exponentially with $l$. One may think that
this last problem can be circumvented by standard purification schemes
\cite{bennett96a,deutsch96,gisin96}. However, purification schemes
require a certain minimum fidelity $F_{\rm min}$ to operate, which cannot be
achieved as $l$ increases. The distance between the nodes is thus
essentially limited by the absorption length of the fiber \cite{footnote}.

In the context of fault-tolerant quantum computing
\cite{fault_tolerant}, using concatenated quantum codes \cite{knill96},
Knill and Laflamme have discussed an important scheme that allows, in
principle, to transmit a qubit over arbitrarily long distances with a
polynomial overhead in the resources. The method requires to encode a
single qubit into an entangled state of a large number of qubits, and to
operate on this code repeatedly during the transmission process. The
tolerable error probabilities for transmission are less than $10^{-2}$,
whereas for local operations they are less than $5\times 10^{-5}$. This
seems to be outside the range of any practical implementation in the
near future. 

In this Letter, we present a model of a {\em quantum repeater} that
allows to create an entangled (EPR) pair over arbitrarily large
distances with a polynomial overhead in the resources and with a
tolerability of errors in the percent region. Once an EPR pair is
created, it can be employed to teleport any quantum information
\cite{bennett93,bennett96b}. Our solution of this problem comprises
three novel elements: $(i)$ a method for creation of entanglement
between particles at distant nodes, which uses auxiliary particles at
intermediate ``connection points'' and a {\em nested purification protocol};
$(ii)$ entanglement purification with imperfect means, including results
for the maximum attainable fidelity $F_{\max}$ and the minimum required
fidelity $F_{\rm min}$; $(iii)$ a protocol for which the time needed for
entanglement creation scales polynomially whereas the required material
resources per connection point grow only logarithmically with the
distance. 

\begin{figure}
\begin{center}
\begin{picture}(230,60)
\put(-8,15){\epsfxsize=130pt\epsffile[0 0 585 72]{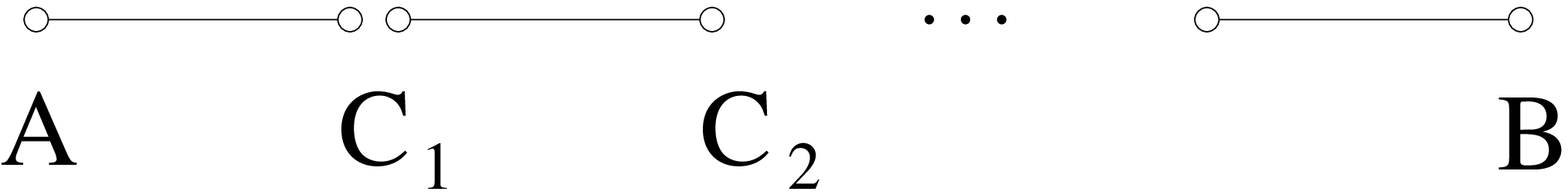}}
\put(150,0){\epsfxsize=85pt\epsffile[0 0 150 92]{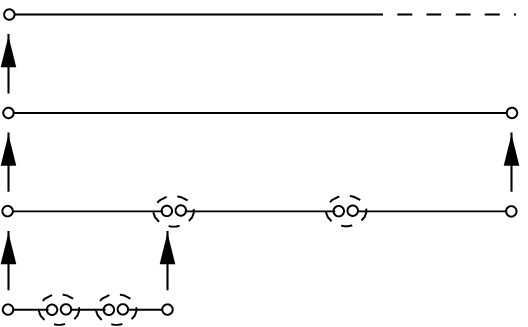}}
\put(-8,50){(a)}
\put(133,50){(b)}
\end{picture}
\end{center}
\caption[]{(a) Connection of a sequence of $N$ EPR pairs. (b) Nested 
purification with repeated creation of auxiliary pairs.}
\label{FIGmanypairs}
\end{figure}

In classical communication, the problem of exponential attenuation 
can be overcome by using repeaters at certain points in the channel,
which amplify the signal and restore it to its original shape. 
Guided by these ideas, for quantum communication, 
we divide the channel into $N$ segments with connection points (i.e. 
auxiliary nodes) in between. We then create $N$ elementary EPR pairs of 
fidelity $F_1$ between the nodes
$A$ \& $ C_1$, $C_{1}$ \& $C_{2}$, \dots $C_{N-1}$ \& $B$, as in
Fig.~\ref{FIGmanypairs}(a). The number $N$ is chosen such that $F_{\rm
min} < F_1\alt F_{\rm max}$. Subsequently, we connect these pairs by 
making Bell measurements at the nodes $C_i$ and classically communicating 
the results between the nodes as in the schemes for teleportation
\cite{bennett93} and entanglement swapping \cite{bennett93,swap}.
Unfortunately, with every connection the fidelity $F'$ of the resulting
pair will decrease: on the one hand, the connection process involves
imperfect operations which introduce noise; on the other hand, even for
perfect connections, the fidelity decreases.
Both effects lead to an
exponential decrease of the fidelity $F_N$ with $N$ of the final pair
shared between $A$ \& $B$. Eventually, the value of $F_N$ drops below
$F_{\rm min}$ and therefore it will not be possible to increase the
fidelity by purification (e.g.\ with the aid of many similar pairs that
are constructed in parallel). The only way to circumvent this limitation
is to connect a smaller number $L\ll N$ of pairs so that 
$F_L > F_{\rm min}$ and purification is possible. 
The idea is then to purify, connect the resulting pairs, purify again, 
and continue in the same vein. The way in which
these alternating sequences of connections and purifications is done
has to be properly designed so that the number of resources needed 
does not grow exponentially with $N$ and thus with the length 
$l$ of the channel ($N\propto l$).

Our proposal, the {\em nested purification protocol}, consists 
of connecting and purifying the pairs simultaneously in the following
sense. For simplicity, assume that 
$N=L^n$ for some integer $n$. On the first level, we
simultaneously connect the pairs (initial fidelity $F_1$) at all the 
checkpoints except at
$C_L,C_{2L},\ldots,C_{N-L}$. As a result, we have $N/L$ pairs of length
$L$ and fidelity $F_L$ between $A$ \& $C_L$, $C_L$ \& $C_{2L}$ and so on. 
To purify these pairs, we need a certain number $M$ of copies 
that we construct in parallel fashion. 
We then use these copies on the segments $A$ \& $C_L$, $C_L$ \& $C_{2L}$
etc., to purify and obtain one pair of fidelity $\ge F_1$ on each segment.
This last condition determines the (average) number of copies $M$ that we
need, which will depend on the
initial fidelity, the degradation of the fidelity under connections, and
the efficiency of the purification protocol. 
The total number of elementary pairs involved in constructing one 
of the more distant pairs of length $L$ is $LM$. 
On the second level, we connect $L$ of these more distant pairs at every 
checkpoint $C_{kL}$ ($k=1,2\ldots$) except
at $C_{L^2},C_{2L^2},\ldots,C_{N-L^2}$. As a
result, we have $N/L^2$ pairs of length $L^2$ between $A$ \& $C_{L^2}$,
$C_{L^2}$ \& $C_{2L^2}$, and so on of fidelity $\ge F_L$. Again, we 
need $M$ parallel copies of these long pairs to repurify up to the fidelity 
$\ge F_1$. The total number of elementary pairs involved in constructing 
one pair of length $L^2$ is thus $(LM)^2$. 
We iterate the procedure to higher and higher levels, until we reach the 
$n$--th level. As a result, we have obtained a final pair between 
$A$ \& $B$ of length $N$ and fidelity $\ge F_1$. 
In this way, the total number $R$ of elementary pairs will be $(LM)^n$. 
We can re-express this result in
the form
\begin{equation}
 R = N^{\log_{L}M+1}
\label{resources}
\end{equation}
which shows that the resources grow polynomially with the distance $N$.
A similar formula was obtained in \cite{knill96} for the overhead
required in propagating the concatenated quantum code. Note that $R$ depends 
only on $L$ and $M$. In order to evaluate $M$, we need to
know the specific form of the error mechanisms involved in the
purification and connections, which in turn depend on the specific
physical implementation of the quantum network. In general, we have only
limited knowledge of these details. In order to estimate $M$, we will 
choose a generic error model for imperfect operations and measurements. 

We define {\em imperfect operations} on states of one or more qubits
by the following maps
\begin{eqnarray}
 \rho &\longrightarrow & O_1 \rho = p_1 \rho_{\text{ideal}} 
 + \frac{1-p_1}{2}\text{tr}_1\{\rho \} \otimes  I_1
\label{one_qubit_op}\\
 \rho &\longrightarrow & O_{12} \rho = p_2 \rho_{\text{ideal}} 
 + \frac{1-p_2}{4}\text{tr}_{12}\{\rho\} \otimes  I_{12}\,,
\label{two_qubit_op}
\end{eqnarray}  
the first of which describes an imperfect one-qubit operation on particle 1, 
and the second an imperfect two-qubit operation on particles 1 and 2.
In these expressions, $\rho_{\text{ideal}}$ is the state that results 
from an {\em ideal} operation, and $I_1$ and $I_{12}$ denote
unit operators on the subspace where the ideal operation acts. The
quantities $p_1$ and $p_2$ measure the {\em reliability} of the
operations. The expressions (\ref{one_qubit_op}) and
(\ref{two_qubit_op}) describe a situation where we have no knowledge
about the result of an error occuring during some operation (``white
noise''), except that it happens with a certain probability $(1-p_j)$. 
An {\em imperfect measurement} on a single qubit in the
computational basis is described by a POVM corresponding to 
\begin{eqnarray}
 P_0^{\eta} &=& \eta |0\rangle\langle 0| + (1-\eta ) |1 \rangle\langle 1|\,, 
\nonumber \\
 P_1^{\eta} &=& \eta |1\rangle\langle 1| + (1-\eta ) |0 \rangle\langle 0|\,.
\label{povm}
\end{eqnarray}
The parameter $\eta $ is a measure for the quality of the projection onto 
the basis states. For example, for the state 
$\rho =|0\rangle\langle 0|$
the measuring apparatus will give the wrong result (``1'') with probability 
$1-\eta \ge 0$. A detailed discussion of this and more general models for
imperfect operations will be given elsewhere \cite{giedke98}. With these
error models we have a toolbox to analyze all the processes involved in the
connection and purification procedures. For example, 
the Bell measurement required in the connection can be decomposed into a 
controlled-NOT (CNOT) operation, effecting e.g.\ 
$|0\rangle|0\rangle\pm|1\rangle|1\rangle \rightarrow 
(|0\rangle\pm|1\rangle)|0\rangle$, followed by two single-qubit measurements.

The basic elements of the nested purification protocol are: (i) pair
connections; (ii) purification. In the following we analyze these
elements using the error models introduced above. Assume now that all of 
the pairs in Fig.~\ref{FIGmanypairs}(a) are in Werner states
(which can be achieved using depolarization \cite{bennett96a}).
Connecting $L$ neighboring pairs as explained earlier, one obtains a new 
``$L$-pair'' with fidelity 
\begin{equation}
 F_L = \frac{1}{4}+\frac{3}{4}\left(\frac{p_1p_2(4\eta^2-1)}{3}\right)^{L-1} 
\left(\frac{4F-1}{3} \right)^L\, .  
\label{connect}
\end{equation}
This formula describes an exponential decrease of the resulting fidelity, 
unless both the elementary pairs and all the operations involved in the 
connection process are perfect. There are several possibilities to do the
purification, and we first generalize the scheme introduced by Bennett
{\em et al.} \cite{bennett96a} to the case of imperfect gate
and measurement operations. In short, the scheme takes two adjacent
$L$-pairs of fidelity $F$, performs local (1 \& 2-bit) operations on the
particles at the same ends of the pairs, and obtains with a certain
probability $p_{\rm succ}$ a new pair of fidelity  
\widetext
\begin{equation}
 F' = \frac{[F^2+(\frac{1-F}{3})^2][\eta^2+(1-\eta)^2]
       +[F(\frac{1-F}{3})+(\frac{1-F}{3})^2)][2\eta(1-\eta)]
       +(\frac{1-p_2^2}{8p_2^2})}{[F^2+\frac{2}{3}F(1-F)
       +\frac{5}{9}(1-F)^2][\eta^2+(1-\eta)^2]
       +[F(\frac{1-F}{3})+(\frac{1-F}{3})^2)][8\eta(1-\eta)]
       +4(\frac{1-p_2^2}{8p_2^2})}\,.
\label{modified_bennett}
\end{equation}
\narrowtext\noindent
The value of $p_{\rm succ}$ is given by the denominator of this
expression. For perfect operations, $\eta=1$
and $p_2=1$, (\ref{modified_bennett}) reduces to the formula given in
Ref.~\cite{bennett96a}.

\begin{figure}
\begin{center}
\begin{picture}(190,200)
\put(-40,10){\epsfxsize=190pt\epsffile[0 0 565 731]{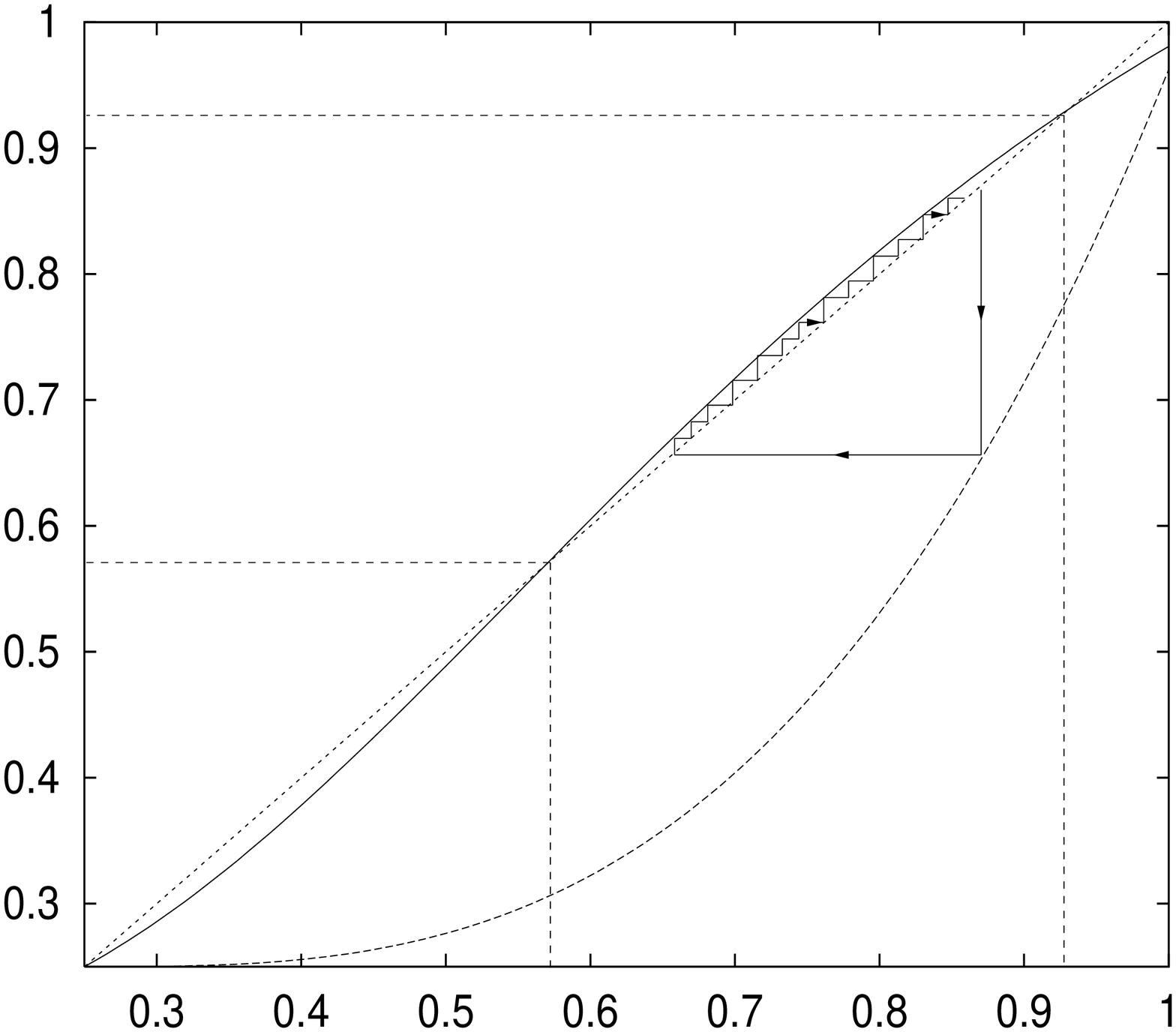}}
\put(97,0){$F$}
\put(120,150){$F'$}
\put(148,84){$F_L$}
\put(10,100){$F_{\text{min}}$}
\put(10,182){$F_{\text{max}}$}
\end{picture}
\end{center}
\caption[]{`Purification loop' for connecting and purifying EPR pairs.
Parameters are $L=3$, $\eta=p_1=1$, and $p_2=0.97$.}
\label{FIGpuriloop}
\end{figure}

Figure \ref{FIGpuriloop} shows the curves for connection (\ref{connect})
and purification (\ref{modified_bennett}) for a certain set of
parameters. The purification curve has two intersection points
with the diagonal, which are the fix points of the map
(\ref{modified_bennett}). The upper point, $F_{\rm max} < 1$
is an attractor and gives the maximum value of the fidelity beyond which 
no pair can be purified. Note also the existence of the minimum value 
$F_{\rm min}>1/2$. 
Together, they define the interval within which purification is possible.
The connection curve, which looks like a simple power in
Fig.~\ref{FIGpuriloop}, stays
below the diagonal for all values of $F$ between $1/4$ and $1$. The
offset of this curve at $F=1$ from the ideal value $F'=1$ quantifies the
amount of noise that is introduced through imperfect operations in the
connection process.

With the above results, we can now analyze the nested purification
protocol. Let us consider a given level $k$ in this protocol, where we
have $N/L^{k-1}$ pairs of fidelity $F$ each. The two-step process
connection--purification can now be visualized as follows (see
Fig.~\ref{FIGpuriloop}). Starting from $F$, the fidelity
$F_L$ after connecting $L$ pairs can be read off from the curve below
the diagonal. Reflecting this value back to the diagonal line, as
indicated by the arrows in Fig.~\ref{FIGpuriloop}, sets the starting
value for the purification curve. If $F_L$ lies within the purification
interval, then iterated application of (\ref{modified_bennett}) leads back 
to the initial value $F$ (staircase). Once the initial value $F$ is
reobtained, we have $N/L^k$ pairs and we can start with the level $k+1$.
In summary, each level in the protocol corresponds to one cycle in
Fig.~\ref{FIGpuriloop}. Note that if, in the loop, $F_L\le F_{\rm min}$
then purification is not possible. Being polynomial in $L$, the lower
curve gets steeper and steeper near $F=1$ for higher values of $L$. From
this, one sees that for a given starting fidelity $F$, there is a
maximum number of pairs one can connect before purification becomes
impossible.

For the resources we obtain $M=\prod_m^{m_{\text{max}}}
2/p^{(m)}_{\text{succ}}$ where $p^{(m)}_{\text{succ}}$ is the 
probability for increasing the fidelity in the $m$-th 
purification step. The total number of steps, $m_{\text{max}}$, is the 
same as in the staircase of Fig.~\ref{FIGpuriloop}.

\begin{figure}
\begin{center}
\begin{picture}(230,90)
\put(0,10){\epsfxsize=110pt\epsffile[31 184 553 603]{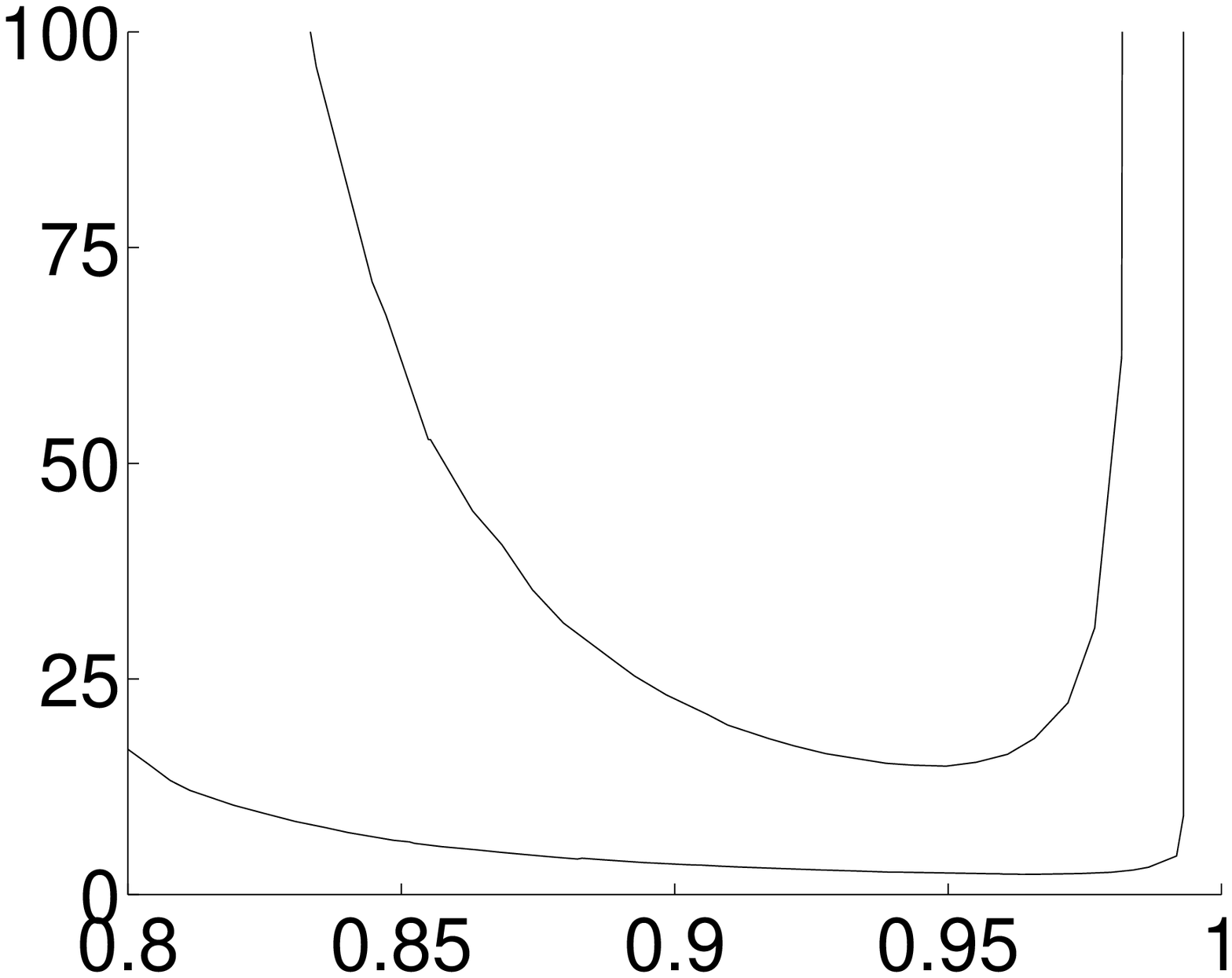}}
\put(123,10){\epsfxsize=110pt\epsffile[64 198 549 596]{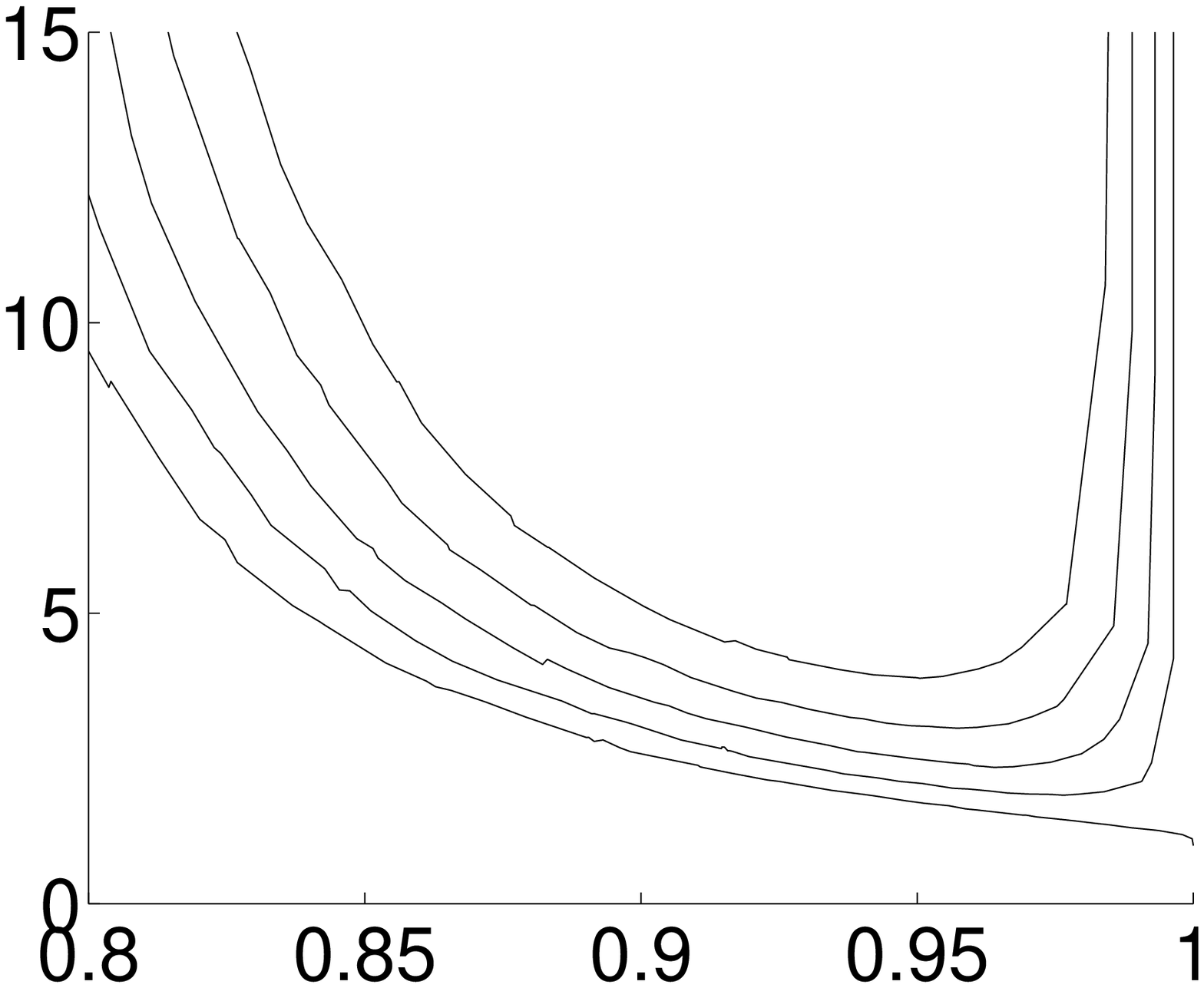}}
\put(-7,53){$M$}
\put(113,53){$M$}    
\put(176,-3){$F$}
\put(57,0){$F$}
\put(57,80){(a)}
\put(176,83){(b)}
\end{picture}
\end{center}
\caption[]{$M$ (see text) versus working fidelity $F$. 
(a) Realization of the repeater with the aid of the purification schemes 
of Refs.~\cite{bennett96a} (upper curve) and \cite{deutsch96} (lower curve). 
The error probabilities of all operations are 0.5\% (error parameters 0.995), 
and $L=2$. (b) Lower curve in (a) for different error probabilities. 
From bottom to top: $0\%$, $0.25\%$, $0.5\%$, $0.75\%$, $1\%$.}
\label{FIGbennett}
\end{figure}

In Fig.~\ref{FIGbennett}(a), $M$ is plotted against the working 
fidelity $F$. Due to the discrete nature of the purification process, 
the fidelity of the repurified pairs need not be exactly the same on each 
nesting level. The working fidelity is thus 
defined as the fidelity maintained {\em on average} when going through 
different nesting levels. The error parameters for this plot are 
$\eta=p_1=p_2=0.995$. One can see that there exists an optimum working 
fidelity of about $0.94$ which requires a minimum number of about 
15 resources. 

A purification protocol that converges faster and therefore involves
less parallel channels was proposed by Deutsch {\em et al.}
\cite{deutsch96}. We have employed this protocol, using imperfect
operations (\ref{one_qubit_op})--(\ref{povm}). As is demonstrated in
Fig.~\ref{FIGbennett}(a), $M$ can be reduced by a factor of the order of
10. Since this number has to be taken to the $n$th power, this reduces
the number of total resources by many orders of magnitude, as is
discussed in Table~\ref{TABresources}. In Fig.~\ref{FIGbennett}(b), $M$
is plotted versus the working fidelity for different error parameters.
One can see that for errors in the one-per-cent region, a working fidelity
can be maintained with on average 5 $L$-pairs on each nesting level. We
note that the procedure also works for error probabilities up to about
$3\%$, but the number of purification resources gets larger.

In the remainder of this paper we propose a protocol for which the resources
grow only logarithmically with the distance, whereas the total {\em
time} needed for building the pair scales polynomially. Imagine that we
purify a pair not with the help of $M$ copies, but instead with one
auxiliary pair of constant fidelity $\pi_0$ that is repeatedly created
at each purification step. The purification with the help of such a pair
leads to a maximum achievable fidelity $F_{\text{max}}(\pi_0)$ that
depends on the value of $\pi_0$ and, more generally, on the state of the
auxiliary pair. This purification method is a variant 
of the standard schemes \cite{bennett96a,deutsch96}, 
with the important difference that the purification limit $F_{\text{max}}$ 
for this method is usually smaller than for the destillation method.  
In the context of the repeater protocol, it is therefore not a priori clear 
whether the fidelity that is lost by the connection process can be regained
with this variant of the purification method.

When connecting $L$ pairs of fidelity $F$ as in Fig.~\ref{FIGmanypairs}(b),
we obtain a resulting $L$-pair of fidelity $\pi_0\equiv F_L$. In the
first step, this pair is swapped to two auxiliary particles at the ends
of the $L$-pair, as indicated by the arrows in
Fig.~\ref{FIGmanypairs}(b). In the next step, an $L$-pair of fidelity
$\pi_0$ is again created by using the same string of particles as
before, which is now used to purify the pair stored between the
auxiliary particles. This procedure can be iterated and thus the stored
pair be purified back to the fidelity $F$ {\em given that} the
purification condition $F_{\text{max}}(F_L) > F$ is satisfied. If this
is the case, then the same procedure can be applied at higher levels,
thereby purifying correlations between more and more distant particles
as indicated in Fig.~\ref{FIGmanypairs}(b). We find that the scheme of
Ref.~\cite{bennett96a} does not satisfy this condition, whereas the 
scheme of Ref.~\cite{deutsch96} generally does. 

In Table~\ref{TABresources}, the total time $T$ and the resources $M^n$
needed to maintain (or distribute) a fidelity of 96\% over a typical
``continental'' (1280km) and ``intercontinental'' (10240km) distance are 
listed. We compare three situations, when the purification part of the 
repeater protocol is realized by (A) the scheme of Bennett {\em et al.}
\cite{bennett96a}, (B) the scheme of Deutsch {\em et al.} \cite{deutsch96},
and (C) using an auxiliary pair of constant fidelity as described above. 
In calculating $T$, two time scales enter: The time $\tau_{\rm op}$
needed for a local operation and measurement, and the time $\tau_{\rm comm}$
needed to communicate measurement results between the nodes. The time for 
creating an elementary pair depends both on $\tau_{\rm op}$ and  
$\tau_{\rm comm}$, and on the specific physical implementation \cite{duer98}.
To estimate orders of magnitude, we assume that $\tau_{\rm op}=10^{-5}s$
and that the repeaters are placed at distances of 10km, corresponding to 
the absorption length of standard optical fibers.  
For the time needed to create an elementary pair we have used the model of 
the photonic channel \cite{vanenk97b,vanenk97a} which gives a typical 
value of $3\times 10^{-4}$s.
These results demonstrate that our scheme for a quantum repeater
allows quantum communication over distances much longer than the absorption 
length.

This work was supported in part by the Austrian Science Foundation, and
by the TMR network ERB-FMRX-CT96-0087. 



\twocolumn
\input{epsf}

\begin{table}
\begin{center} 
\begin{tabular}{|l|c|c||c|c|} 
 & \multicolumn{2}{c||}{Continental scale} & 
\multicolumn{2}{c|}{Intercontinental scale} \\ \hline 
 & resources & time $[s]$ & resources & time $[s]$  \\ \hline
   A& $1.58*10^9$ & $3.88*10^{-2}$ & $9.01*10^{12}$ & 0.298 \\ \hline
   B & 329 &  $1.34*10^{-2}$ & 4118 & 0.103 \\ \hline
   C & 7 &  0.241 & 10 &3.275\\ 
\end{tabular}
\end{center}
\caption[]{Parallel resources $M^n$ and time $T$ needed for creating a 
distant EPR pair via optical fibers (see text).  
Continental scale means $2^7=128$ segments, intercontinental  
scale means $2^{10}=1024$ segments. Error 
parameters are $\eta=p_1=p_2=0.995$. For (C), the resources grow only 
logarithmically, i.e. $M^n=n+1$.}
\label{TABresources}
\end{table}


\begin{references}
\bibitem{footnote}
    For standard optical fibers, this length is typically 10km 
    (see \cite{tittel97}).
\end{references}
\end{document}